\documentclass[%
 twocolumn,reprint,
 amsmath,amssymb,
 aps,
]{revtex4-2}

\usepackage{graphicx}
\usepackage{dcolumn}
\usepackage{bm}

\begin{document}

\preprint{APS/123-QED}

\title{Statistical Dynamics of Wealth Inequality \\
in Stochastic Models of Growth \\
 }

\author{Jordan T. Kemp$^1$ and Lu\'is M. A. Bettencourt$^{2,3}$}
 \affiliation{%
 $^1$Department of Physics, University of Chicago, Chicago, Illinois 60637, USA}%
\author{}%
\affiliation{%
 $^2$Department of Ecology and Evolution, University of Chicago, Chicago, Illinois 60637, USA}%
\affiliation{%
 $^3$Mansueto Institute for Urban Innovation, University of Chicago, Chicago, Illinois 60637, USA}%

\date{\today}

\begin{abstract}
Understanding the statistical dynamics of growth and inequality is a fundamental challenge to ecology and society. 
Recent analyses of wealth and income dynamics in contemporary societies show that economic inequality is very dynamic and that individuals experience substantially different growth rates over time. However, despite a fast growing body of evidence for the importance of fluctuations, we still lack a general statistical theory for understanding the dynamical effects of heterogeneneous growth across a population.  Here we derive the statistical dynamics of correlated growth rates in heterogeneous populations. We show that correlations between growth rate fluctuations at the individual level influence aggregate population growth, while only driving inequality on short time scales. We also find that growth rate fluctuations are a much stronger driver of long-term inequality than earnings volatility. 
Our findings show that the dynamical effects of statistical fluctuations in growth rates are critical for understanding the emergence of inequality over time and motivate a greater focus on the properties and endogenous origins of growth rates in stochastic environments.
\end{abstract}

\maketitle
Exponential growth and broad inequality are general features of population dynamics in biology and society~\cite{scheffer}. In biology, the maintenance of diverse populations is associated with greater biodiversity and with larger pools of variability enabling faster processes of evolution by natural selection~\cite{frank}. In human societies, rates of long-term (economic) growth are much higher than in most ecosystems, often generating widening inequalities and leading to familiar stresses of social justice and equity~\cite{piketty,flavio}. 

Researchers have recently taken a special interest in the dynamics of economic inequality~\cite{scheffer}, applying multidisciplinary approaches that include historical data analyses~\cite{keister2000wealth} and searches for social, political and economic mechanisms that generate, distribute, and rebalance wealth within populations~\cite{piketty,during2008kinetic}. 
Research in economics has emphasized the importance of heterogeneity in populations~\cite{flavio} as a source of widening wealth inequality. These heterogeneities associated with a number of different population features such as the divergence of incomes between capital and labor~\cite{piketty}, between management and workers within firms~\cite{bakija,atkinson}, and between people with different educational attainment~\cite{goldin}. These primarily empirical approaches point to the fundamental importance of diverging income growth for different subpopulations and point to the need for rebalancing mechanisms if inequality is to remain controlled. 
However, we still do not have statistical theories applicable in general circumstances and, as a consequence, we lack a clear picture of how to concurrently manage growth and inequality.  

To this end, research in physics has derived a number of results on the role of stochasticity as a source of inequality, on inter-agent exchanges (particularly in the mean-field limit), and proposed redistribution mechanisms in the context of stochastic geometric growth models~\cite{bouchaud1, garlaschelli, degond, medo, chakraborti, patriarca}. The main focus of this work has been the design of wealth redistribution schemes towards creating long time stationary limits~\cite{bouchaud,berman2, li} with parametrically controlled levels of inequality and whether such stationary solutions exist~\cite{peters, berman}. 

While these models have been used to fit data on wealth distributions \cite{li,berman}, their specialized approaches have left open questions on the fundamental statistical dynamics underpinning the generation and allocation of wealth. For example, evidence suggests models following strict application of Gibrat's law (individual growth rates independent of scale) cannot characterize the rapid emergence of inequality experienced in recent years~\cite{gabaix, garlaschelli}, motivating explicit analysis of open-ended dynamical effects due to various sources of fluctuations. 
These developments support multiplicative stochastic growth as a starting point for modeling~\cite{bettencourt}, but call for the consideration of statistical effects due to fluctuations and correlations in model parameters, especially growth rates.

To address these issues, we derive the time evolution of the resource (wealth) distribution in a statistical population of heterogeneous agents experiencing geometric random growth. We focus on the dynamical effects of growth rate fluctuations and the effects of their correlations with resources across agents and over time.
We show that these two effects lead to two dynamical time scales, which require different control measures so that inequality does not explode in a population over time. Specifically, we show that natural schemes to reduce inequality are subject to reversal over longer time scales because of variability of growth rates in the population. We end by discussing the mechanisms that may simultaneously lead to sustained exponential growth and control of long term sources of inequality in terms of agent-based learning in correlated stochastic environments.

The fundamental model for the dynamics of resources in populations relies on stochastic exponential growth \cite{gabaix,bettencourt}. In its simplest form, known as geometric Brownian motion, agents generate wealth by (re)investing incomes net of costs. Crucially these models, unlike additive stochastic growth, generate population lognormal and power-law statistics, which characterize observed wealth and income statistics \cite{zipfs,clauset} 

Specifically, the dynamics of stochastic multiplicative growth start by tracking the resources (wealth), $r(t)$, of a specific agent at time $t$. This quantity changes in time via the difference between an income, $y(t)$, minus costs, $c(t)$. The difference, $y-c$ (net income), is then defined to be proportional to resources, expressed as a (potentially $r$ dependent) stochastic growth rate, $\eta(t)=(y(t)-c(t))/r(t)$, for $r>0$.
It follows that the time evolution of resources obeys a simple (but multiplicative) stochastic differential equation $dr(t) = \eta r dt + \sigma r dW_t$, where the volatility, $\sigma$, is the standard deviation of the average growth rate, $\eta$.
When these parameters are independent of $r$ and $t$, this stochastic equation can be integrated via It$\bar{\textrm{o}}$ calculus \cite{bettencourt} to give
\begin{equation}\label{res}
    \ln\frac{r(t)}{r(0)}=\bigg({\eta}-\frac{\sigma^2}{2}\bigg)t+\sigma W(t),
\end{equation}
where $\gamma={\eta} - \sigma^2/2$ is the effective growth rate and $r_0$ are the agent's resources at $t=0$, $r_0=r(0)$. The decrease in the mean growth rate $ {\eta} $ due to finite volatility is an important feature of multiplicative growth. The quantity $\sigma W(t)$ is a Wiener process with magnitude proportional to the volatility and units of $t^{-1/2}$. 

Under these circumstances the population dynamics follow from the single agent time evolution, but correlations between $\eta$ and $r_0$ introduce new considerations for the average growth rate. When growth rates are identical and statistically independent in the population, the average growth rate is the average over the growth rates of all individuals. However, when initial log wealth $\ln r$ and growth rates $\eta$ are correlated, the growth of the average becomes (Appendix \ref{apA}) $(\gamma r)_N=\big(1+\textrm{covar}_N(\frac{\gamma_i}{\bar\gamma},\frac{ r_i}{\bar r})\big)\bar\gamma\bar r$
for population average resources $\bar r$,  and covariance in agent parameters over the population $\textrm{covar}_N$. 

\begin{figure}
    \centering
   \hspace*{-.8cm}\includegraphics[width=.57\textwidth]{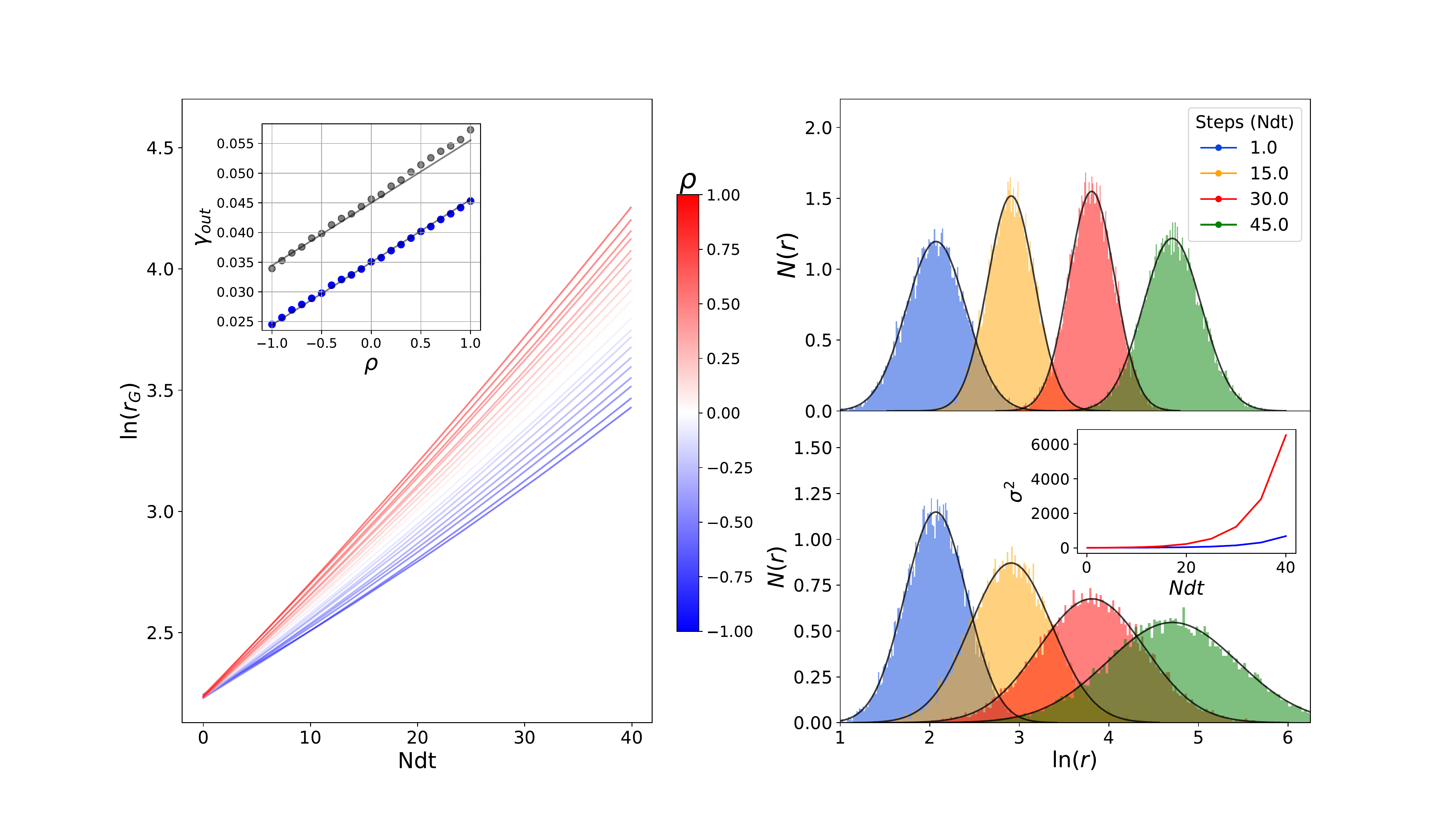}
    \caption{ Monte Carlo simulations of growth dynamics with correlated growth rates and initial resources. \textit{Left:}  Aggregate resource trajectories, of populations with $\rho$ ranging from $-1$ (blue) to $1$ (red). \textit{Inset}: The population averaged effective  growth rate (blue) scales linearly in $\rho$, following Eq.~\ref{covarGen}. The temporal averaged growth rate (gray) scales similarly via $\gamma=\gamma^\prime+\sigma^2$ \cite{bouchaud1}. \textit{Right:} Distributions of log wealth at different time steps for a regressive, $\rho=.75$ (bottom),  and progressive, $\rho=-.75$ (top) population. Black curves represent normal distribution fits. The regressive assignments broaden more quickly, indicating that the populations are becoming more rapidly unequal. \textit{Inset:} Variance of the progressive (blue) and regressive (red) populations over many time steps. Data are simulated from populations of $10^5$ agents with Gaussian distributed initial resources of mean $\bar r=10$, mean growth rate $\eta=.04$, and standard deviations $\sigma_\eta=.015,\sigma_r=.682$. }\label{fig:combined}
\end{figure}

We denote the term, $\gamma^\prime=\bar\gamma+\textrm{covar}_N \big(\gamma_i,\frac{ r_i}{ \bar r}\big)$, the population effective growth rate. This quantity can be calculated analytically for normally distributed quantities $\gamma$ and $\ln r/{\bar r}$. The result involves the variances of both quantities as well as the Pearson correlation coefficient between them, as 
$\textrm{covar}_N(\gamma_i,\frac{r_i}{\bar r}) =\rho\sigma_G \sigma_r e^{\sigma^2_r/2\bar r}$.  The effective growth rate becomes
\begin{equation}\label{covarGen}
\gamma^\prime=\eta-\frac{\sigma^2}{2}+\rho\sigma_G \sigma_r e^{\sigma^2_r/2\bar r},
\end{equation}
where $y\equiv\ln r/\bar r$, and the lognormal distributed resources introduce an exponential term. If instead resources were normally distributed, this term becomes unity. This expression shows that a positive covariance between initial log resources and growth rates results in a higher population effective growth rate and vice versa. Of course, associating higher-growth opportunities with wealthier individuals will exacerbate inequalities in the population over time.  We denote populations with this configuration as \textit{regressive} and with negative correlation as \textit{progressive}. We visualize the tradeoff between short-term change in inequality and growth  through numerical simulations in Fig.~\ref{fig:combined}.   

We see that these simple considerations present an apparent paradox for any attempt to simultaneously maximize total wealth (social welfare) and reduce inequality. Decreasing $\rho$, thereby making the society more {\it progressive} results in a social opportunity cost in terms of a decline in average growth. This effect may even lead to decline (negative exponential growth) in societies starting out with low average growth and high volatility. One way out of this dilemma is for a progressive policy assignment to create (in ways to be specified) higher average growth rates than the regressive case.

\begin{figure}[h]
\vspace{-.4cm}
\includegraphics[width=.24\textwidth]{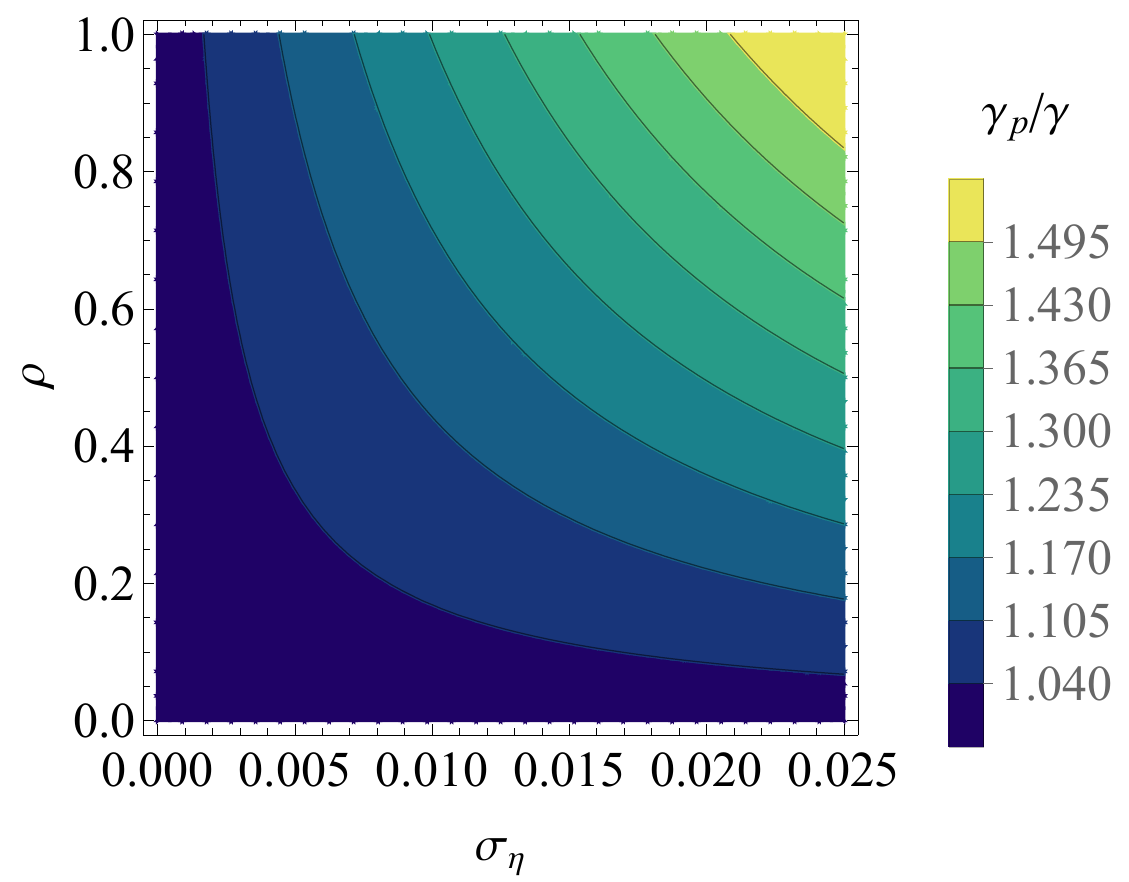}
    \qquad
    \hspace{-.5cm}
    \includegraphics[width=.22\textwidth]{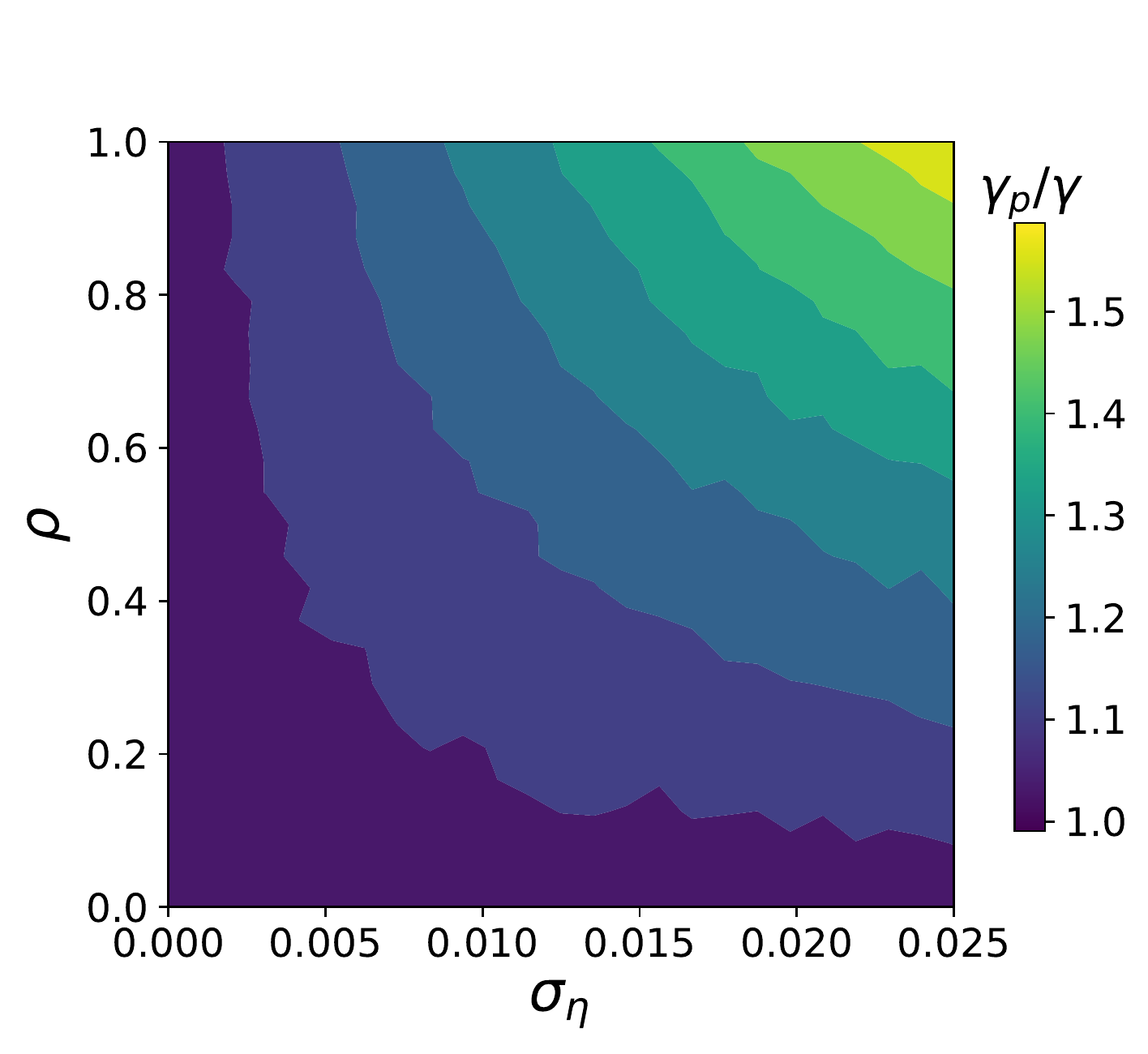}
    \includegraphics[width=.4\textwidth]{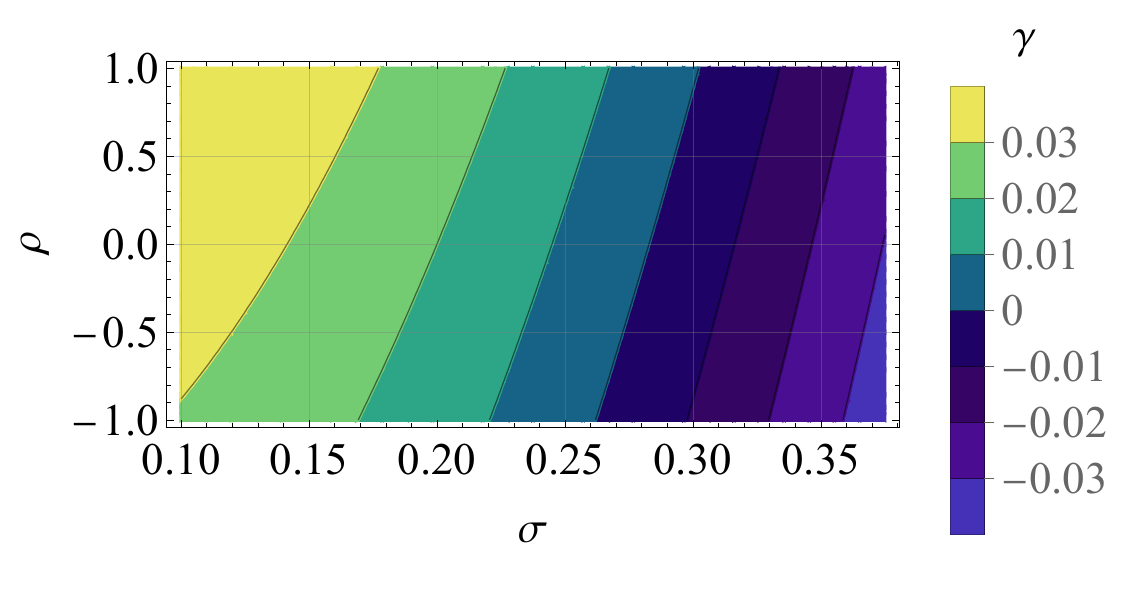}
\vspace{-.5cm}
\caption{\label{fig:HeatMap} Parameter spaces for progressive/regressive growth rate ratios across variances (top) and growth rate regimes across correlation and volatility values (bottom) \textit{Upper Left:} Contour plot of $\gamma_p/\gamma$ for correlations and growth rate variances with condition $\sigma_G= \sigma_G^{(p)}$. \textit{Upper Right:} A Monte-Carlo simulation comparing the correlated average growth rates of populations with identical parameters up to a change in sign of $\rho$ agrees with theory. To stay competitive, the progressive society requires a larger growth rate as either parameter increases.   \textit{Lower:} The effective growth rate for $\eta=.04,\sigma_r=1,\sigma_G=.025$  transitions from positive to negative along the correlation axis at high volatility ($\sigma\approx.275$).}
\end{figure}

To analyze this possibility, Eq.~\ref{covarGen} introduces the threshold correlation below which aggregate effective growth rate becomes negative. Computed under the condition  $\gamma^\prime=0$, it is $\rho_c=r\big(\frac{\sigma^2}{2}-\eta\big) /2\sigma_r\sigma_G$. Similarly, the critical volatility marking the crossover from positive to negative average growth, denoted $\sigma_c$, can be determined by rearranging the expression for $\sigma$. This relationship is plotted in Fig. \ref{fig:HeatMap}. We can directly compare the efficacy of progressive and regressive assignments by computing the ratio of growth rates between a population with growth rate $\gamma$, and its hypothetical progressive counterpart, $\gamma_p$ with distribution parameters with superscript ${(p)}$.  In the simplest case where we assume identical initial population conditions up to the assignment of growth rates (sign of $\rho$), Eq.~\ref{covarGen} produces the condition where average growth is preserved in the progressive society
\begin{equation}\label{etaprime}
\gamma_p= \gamma+\rho\sigma_r\big(\sigma_G+ {\sigma_G}^{(p)}\big)e^{\sigma^2_r/2\bar r}.
\end{equation}
Thus, a progressive arrangement must have a larger average growth rate in order to achieve a population effective growth rate equal to its regressive counterpart. This difference is made worse in regimes with stronger starting inequality and growth rate fluctuations.  Fig. \ref{fig:HeatMap}  demonstrates this equation's agreement with population Monte-Carlo simulations in which $r_0$ is normally distributed.

So far, we have shown that the population average growth rate collects a correction from covariances between resources and growth rate variations. This introduces a trade-off between managing inequality in the short run and maximizing overall growth which can only be resolved, at each time interval, if progressive assignments also result in higher average growth rates.  

But what happens over the longer term, as growth rate fluctuations persist? To answer this question, we consider the solution to the Fokker-Planck equation (FPE) for the time dependent probability of  resources $r$ for each agent.
 The solution with initial conditions $r(0)=r_0$ (a delta function at the individual level) is a lognormal distribution \cite{Stoj} with time dependent mean and variance. This can be re-written as a Gaussian distribution of $\ln r$ as
\begin{equation}\label{lognorm}
    P(\ln r,t|\gamma,\ln r_0)=\frac{d\ln r}{\sqrt{2\pi\sigma^2 t}}\exp\bigg[-\frac{(\ln r-\ln r_0-\gamma t)^2 }{2\sigma^2t}\bigg],
\end{equation}
where at early times, the expected resource value is given by the linear equation $\langle \ln r(t) \rangle= \ln r_0+\gamma t$,  and asymptotically approaches $\gamma t$. The variance increases linearly in time $\langle \delta \ln r \rangle^2=\sigma^2 t$, which shows that individual wealth also becomes more uncertain as a function of time. The critical volatility at which growth ceases is, as before, given by $\sigma_c=\sqrt{2{\eta}}$ . We consider Eqn. \ref{lognorm} the \textit{homogeneous} population solution.

Changing variables and working explicitly with time average growth rates yields further insight into the dynamics. Dividing each factor of the exponential by t, and performing the variable transformation $\frac{1}{t}\ln\frac{r}{r_0}\rightarrow\phi,td\phi=d\ln r$ yields the distribution of growth rates
\begin{equation}\label{dist}
    P(\phi,t|\gamma,\sigma)=\frac{d \phi}{\sqrt{2\pi\sigma^2/t}}\exp\bigg[-\frac{(\phi-\gamma)^2 }{2\sigma^2/t}\bigg].
\end{equation}
This shows explicitly that the statistics of the time averaged growth rates are simpler than those of resources. The average growth rate becomes stationary at long times as the mean remains constant, while growth rate fluctuations decrease in time as $1/t$, converging in distribution to a delta function on a time scale $t >> \sigma^2$. 

This suggests a simple picture emerging at the population level; however varying the growth rates among agents in the population will introduce complications.  The most direct route to assess these effects follows by positing Gaussian distributions on growth rates, from the asymptotic behavior of growth rate distributions \cite{takayasu, wyart, bettencourt}, and initial log resources across the population, from the solution to the FPE.

To obtain the dynamical solution for the distribution of resources in the population, we then convolve Eq.~\ref{lognorm} with a bivariate static Gaussian distribution of mean initial log resources $\mu_0$, variance $\sigma_r^2$, and mean growth rate $G$, variance $\sigma_G^2$ via  (Appendix \ref{apB}) 
\begin{equation}\label{conv}
    P(\ln r,t)= \frac{d\ln r}{\sqrt{2\pi \Sigma^2}}\exp\bigg[-\frac{(\ln r-\mu_0-G t)^2}{2\Sigma^2}\bigg],
\end{equation}
where the variance in the population, $\Sigma^2$, is given by
\begin{equation}\label{var}
    \Sigma^2=\sigma_r^2+(\sigma^2+2\rho\sigma_r\sigma_G) t+\sigma_G^2t^2.
\end{equation}
We consider Eq. \ref{conv} the \textit{heterogeneous} population solution. The quadratic expression in time for variance is characterized by two different timescales, $t_{c1}=\sigma_r^2/(\sigma^2+2\rho\sigma_r\sigma_G)$ and $\ t_{c2}=(\sigma^2+2\rho\sigma_r\sigma_G)/\sigma_G^2$. The natural timescale in this analysis is years~\cite{piketty}, such that \textit{early} times are on the order of a few years, \textit{intermediate} over a few decades, and \textit{long} timescales are several decades to centuries. Effects over long timescales can thus be thought of as generational, although we do not incorporate life cycle dynamics in this analysis. Annual growth rates are typically on the order of a few percent a year ($\gamma \simeq 10^{-2}t^{-1}$), whereas log wealth varies in the population on a typical scale of  $\delta \ln r \simeq 10^1$, so the magnitude of $\sigma_G$ is naturally some orders smaller than $\sigma_r$. This results in $t_{c1}< t_{c2}$, producing three distinct dynamical regimes for the population variance and hence inequality dynamics. For early times $t<t_{c1}$, the variance of resources in the population is given approximately by its initial condition $\Sigma^2\simeq\sigma_r^2$. 
For intermediate times $t_{c1}< t< t_{c2}$,  $\Sigma^2 \simeq (\sigma^2+2\rho\sigma_r\sigma_G) t$.  
The intermediate regime introduces the explicit decrease in inequality from a progressive assignment, so long as  $\sigma^2<2\rho\sigma_r\sigma_G$. 
However, this benefit is short-lived, as for later times $t> t_{c2}$, $\Sigma^2\simeq \sigma_G^2t^2$, and variance invariably explodes due to fluctuations in growth rates. 
Note that the long time regime does not require biases between resources and growth rates; it persists under the weaker conditions of a finite variance in growth rates and appears at earlier times for larger growth rate variances. 

Finally, we observe that the asymptotic population growth rate distribution simplifies to a Gaussian
\begin{equation}\label{resourceLim}
    \lim_{t\rightarrow\infty}P(\phi,t)=\frac{d \phi}{\sqrt{2\pi\sigma_G^2}}\exp\bigg[-\frac{(\phi-G)^2}{2\sigma^2_G}\bigg],
\end{equation}
in contrast to the case of a single agent, reflecting the Gaussian distribution of growth rates in the population.  Over the long term, population dynamics are thus dominated by fluctuations in growth rates, which lead to widening inequality the larger the growth rate variance. 


This effect introduces a general mechanism that can account for increases in inequality that occur over intermediate and long timescales ~\cite{gabaix}, and incorporate findings that empirical growth rates are heterogeneous beyond fluctuation noise, and correlate with wealth over time \cite{guvenen,fagereng2020heterogeneity,hubmer2021sources}. It also asks that we shift our focus from the dynamics of wealth itself to the statistics of its drivers in terms of variations of growth rates. 

We now analyze the detailed dynamics of inequality using two standard metrics. First, the Gini coefficient, $G_{\rm ini}$, is the most common metric measuring inequality within a population. It ranges from $G_{\rm ini}=0$ for equally distributed wealth, and trends towards 1 as a society becomes maximally unequal. For lognormally distributed wealth in the infinite population limit, $G_{ini}=\textrm{Erf}\big[\Sigma/2\big]$.
Second, we examine the social cost of high growth on inequality, a topic of interest and debate in the social sciences~\cite{lucas, cordoba}. Using the coefficient of variation (CV), denoted $c_v$, we compare how quickly a distribution's mean increases relative to its standard deviation \cite{bellu}. The quantity $c_v=\Sigma/(\mu_0+Gt)$ where, from Eq.~\ref{covarGen}, $G=\eta-\frac{\sigma^2}{2}+\rho\sigma_G\sigma_r e^{\sigma_r^2/2\bar r}$. A more equitable economy would decrease $c_v$; growing without increasing inequality. In the egalitarian limit,  $c_v=0$ as every agent has the same resources, and variance is zero. Conversely, an exploding $c_v$ in the positive or negative direction indicates increasing inequality, marked by positive or negative aggregate growth respectively.  From Eq.~\ref{var}, without dynamical redistribution measures, societies will invariably become maximally unequal as multiplicative dynamics are dominated by fluctuations in earnings \cite{scheffer}. 

\begin{figure}[h]
\begin{tabular}{cc}
  \includegraphics[width=.24\textwidth]{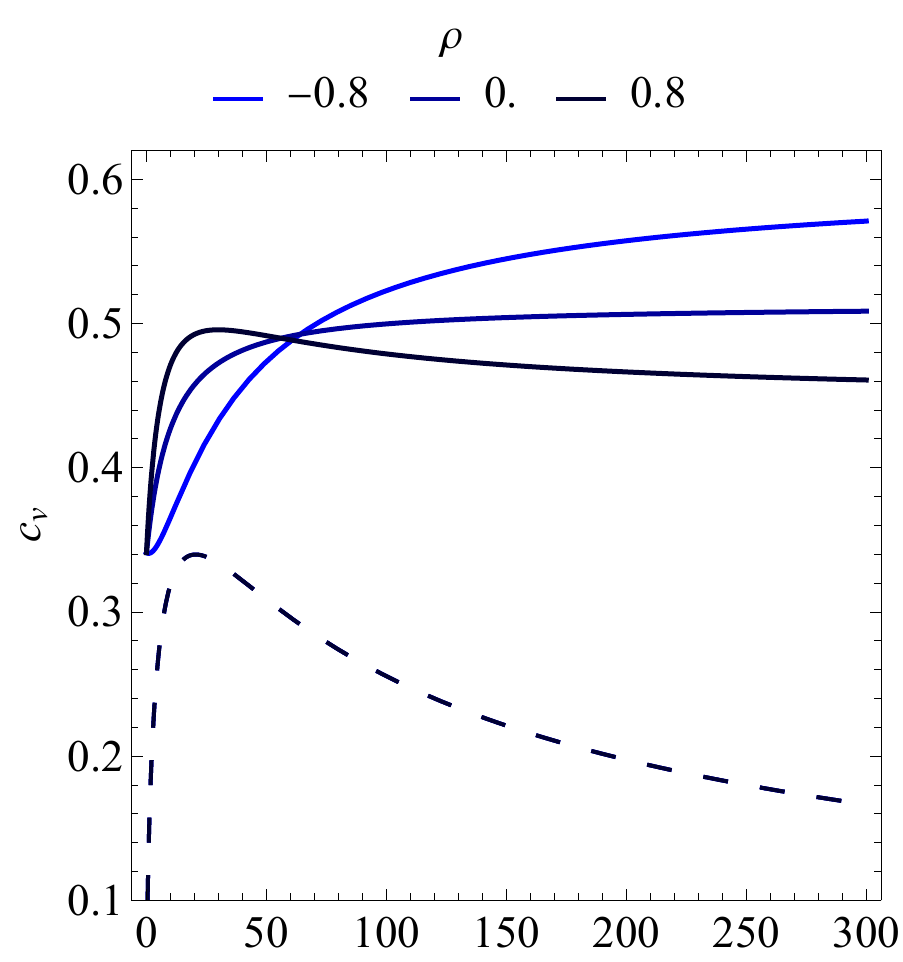} &   \includegraphics[width=.24\textwidth]{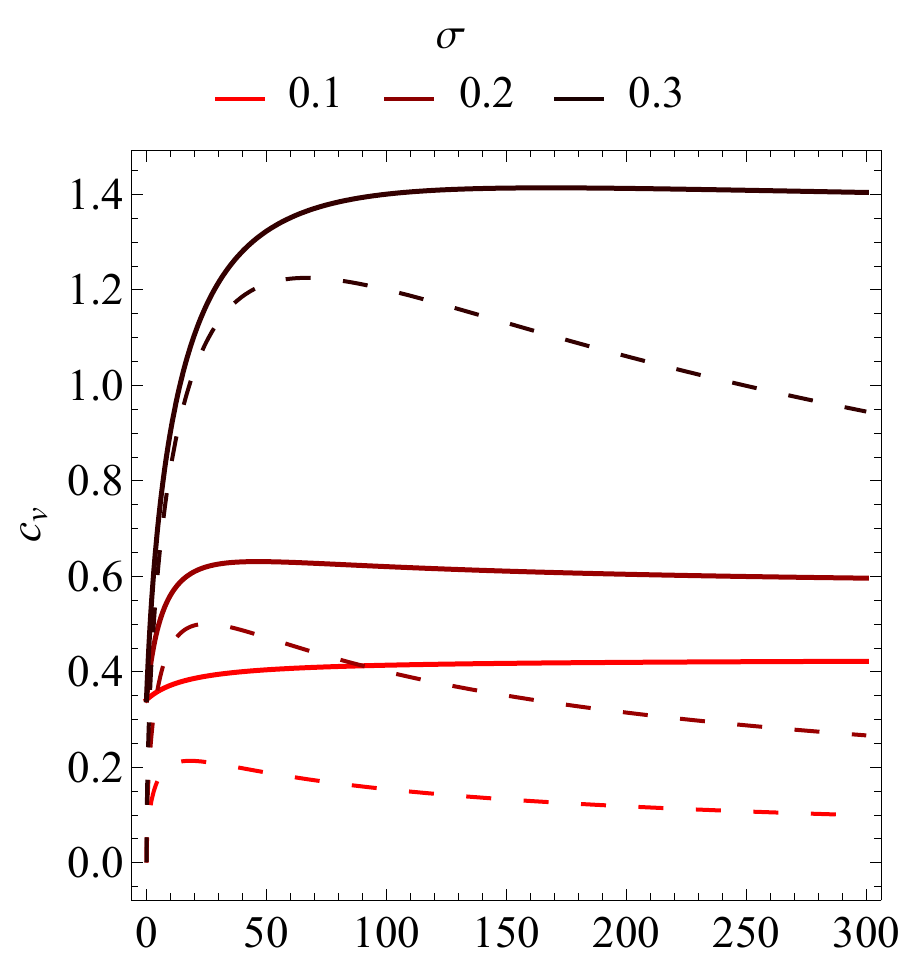} \\
  \includegraphics[width=.24\textwidth]{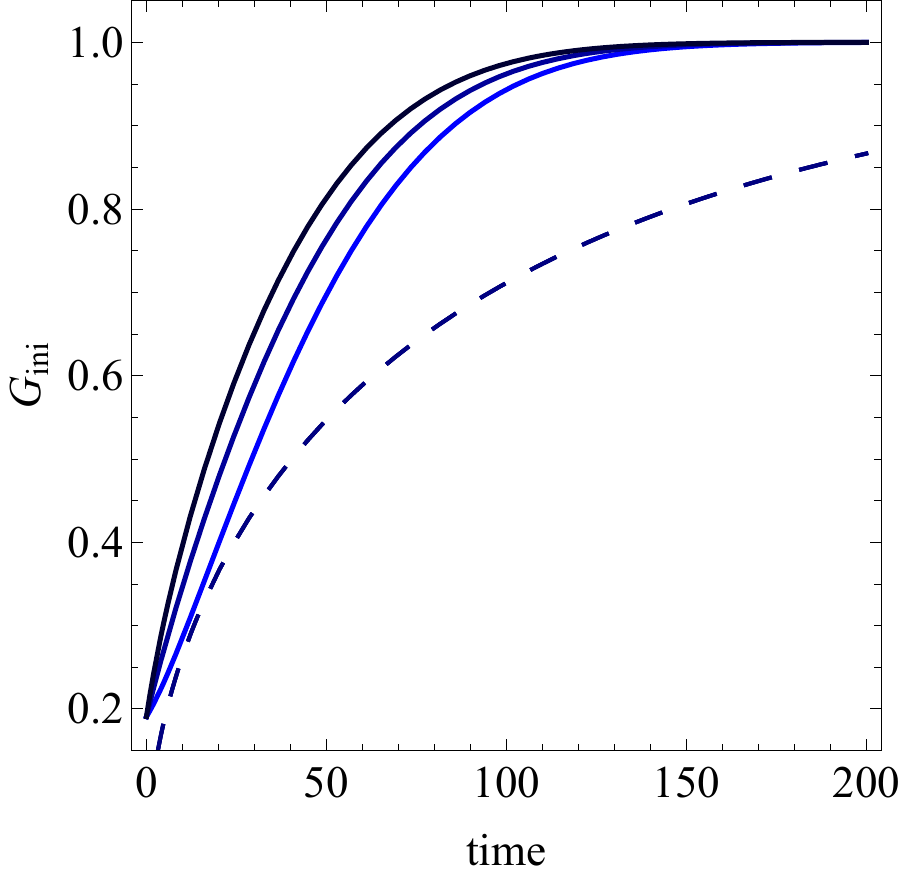} &   \includegraphics[width=.24\textwidth]{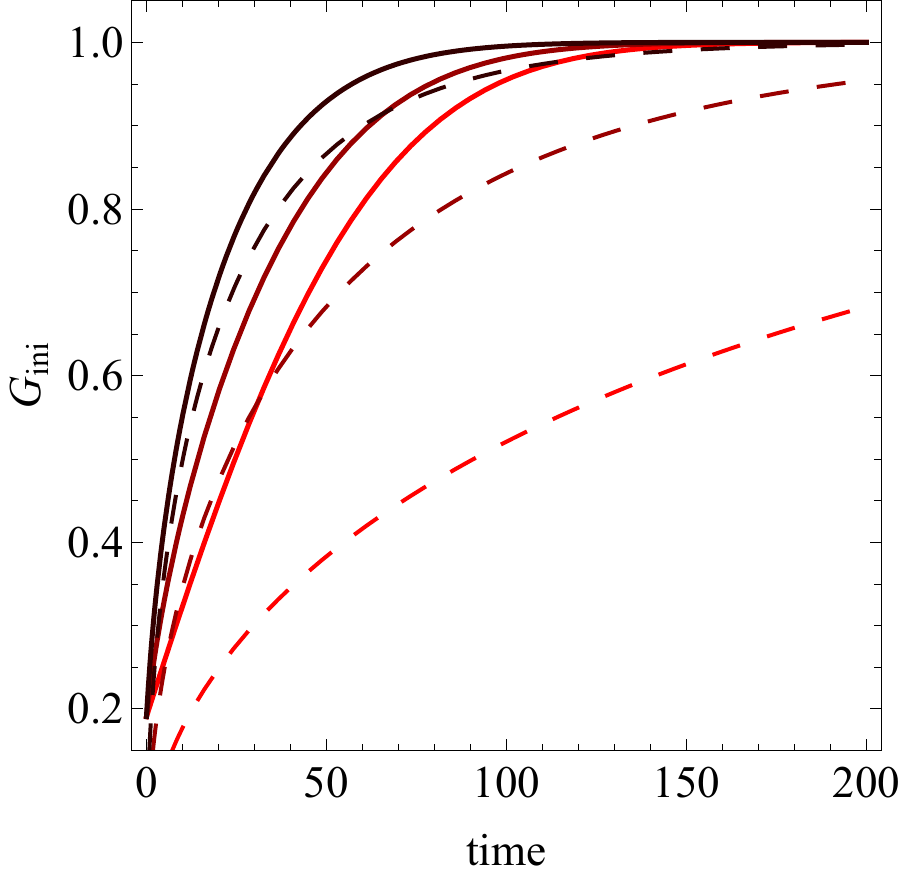} \\
\end{tabular}
\includegraphics[width=.3\textwidth]{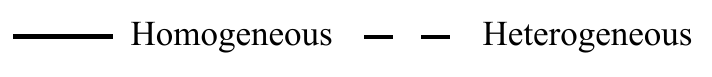} 
\caption{\label{fig:Curves} Trajectories of inequality metrics in various parameter regimes;  $\sigma=.15,\sigma_r=.341,\mu_0=1,\eta=.06,\sigma_G=.025$. See legend for line type mapping.  \textit{Left: }Plots for low volatility, $\sigma=.1$, populations with different correlation coefficient values. At early times, $G_{ini}$ increases more quickly for $\rho<0$, and slightly decreases in a society with low volatility. Variances in growth rates in both cases drive the $G_{ini}$ to 1 more rapidly than in the homogeneous case. For $\rho>0$, $c_v$ jumps higher than in an uncorrelated society, but rapidly approaches a lower asymptotic value due to a higher average growth rate with comparable inequality. Conversely, $c_v$ initially decreases for $\rho<0$, but reaches a higher asymptotic value as long term inequality remains comparable at the cost of a lower average population growth rate. \textit{Right:} Regressive populations ($\rho=.75$) at different values of volatility. A higher volatility causes more rapid increase in $G_{ini}$ in both population configurations, and volatility positively influences the peak and asymptotic values of $c_v$. Progressive assignments hurt long term growth while marginally affecting long-term inequality dynamics, while heterogeneities play a dominant role in accelerating the emergence of  inequality.
}
\end{figure}

\begin{figure}
Homogeneous \\
\includegraphics[width=.24\textwidth]{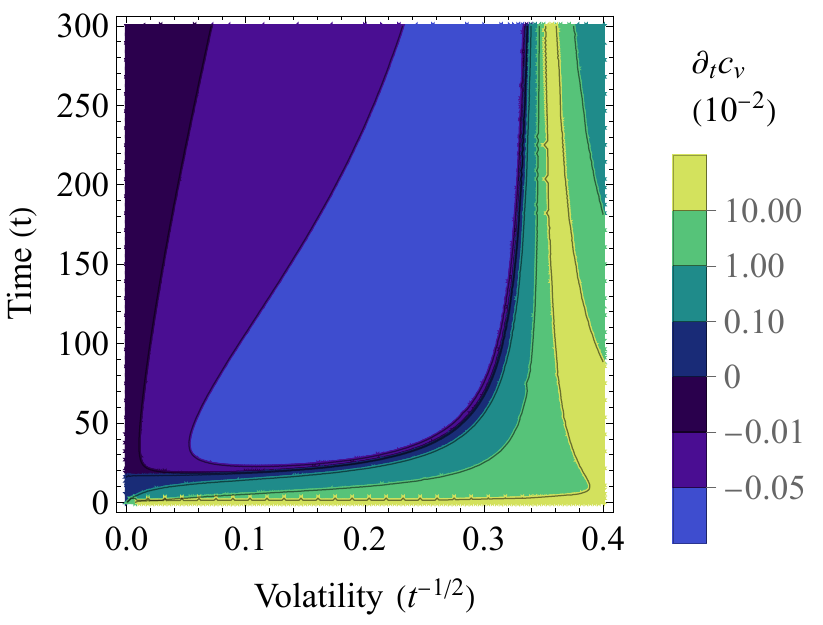}  \\
\begin{tabular}{cc}

Regressive  & Progressive  \\
    \includegraphics[width=.24\textwidth]{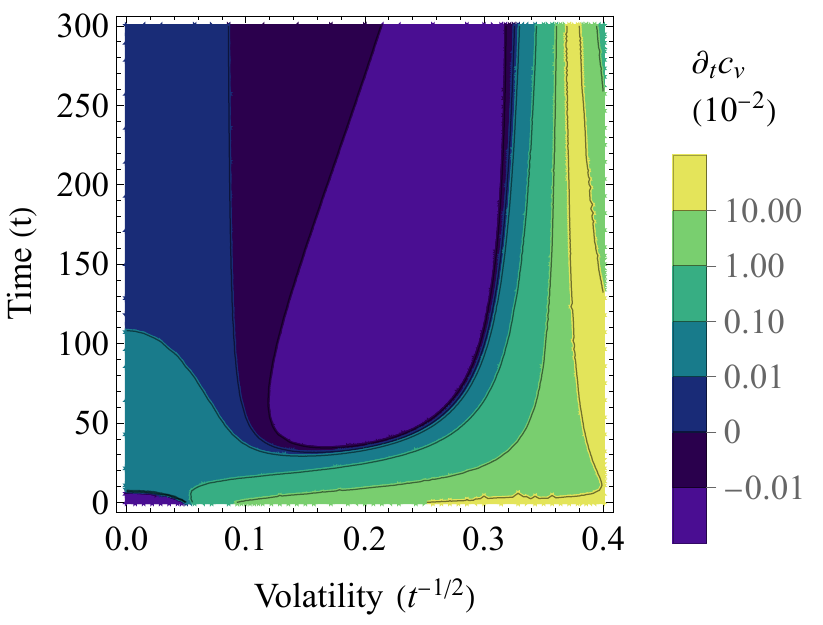}&   \includegraphics[width=.24\textwidth]{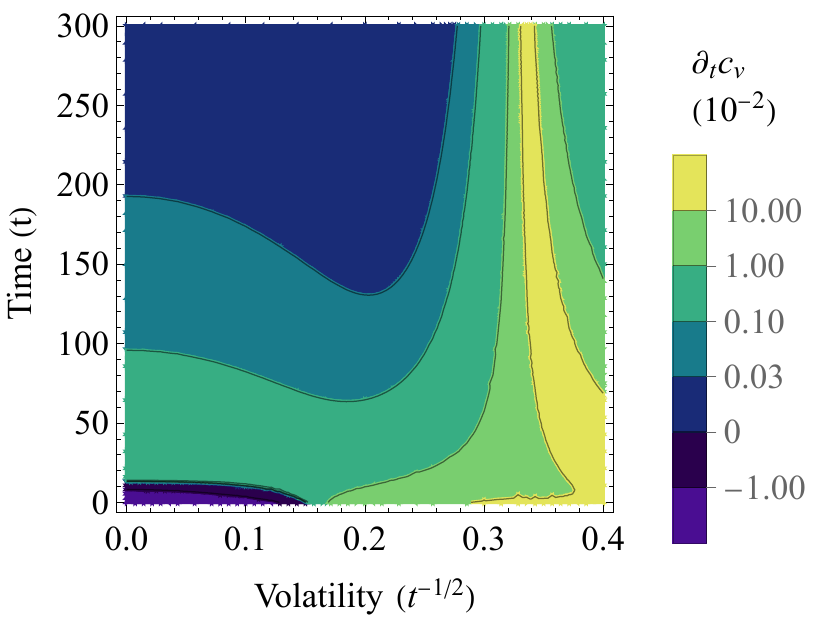} \\

\end{tabular}
\caption{\label{fig:ID}Contour plots for $\partial_t c_v$ in identical populations to Fig. \ref{fig:Curves}, with $\eta=.035$. 
At low volatilities, homogeneous $c_v$ initially increases, then decreases with magnitude increasing in $\sigma$. For $\rho>0$, this behavior persists following initial times of high increase in $c_v$. The initial decrease observed for $\rho<0$ only occurs at low volatility, and $c_v$ does not decrease in later times, as demonstrated in Fig. \ref{fig:Curves}. Past the homogeneous critical volatility, given by $\sigma_c\approx.265$, $c_v$ blows up as the negative growth rate drives average log resources through zero to a negative value, and is seen by the bright yellow bands. More negative $\gamma$, caused by high $\sigma$ or negative $\rho$, reduces the time needed to pass through $\ln r=0$. In this regime of instability, resources continue to decrease as inequality grows. Introducing heterogeneity shifts this crossover point based on the value of the covariance term in the growth rate.}
\end{figure}

We explore the dynamics of homogeneous and heterogeneous populations by comparing time-series trajectories of $c_v$ and $G_{ini}$ across several values of correlation coefficient and volatility. Fig.\ref{fig:Curves}  demonstrates the impact of heterogeneity on inequality (Gini), and its relative effects on growth (CV). In homogeneous populations, volatility spurs inequality from uniform initial conditions causing $c_v$ to initially increase, then peak and decrease as resources increase in $t^{1/2}$ faster than population standard deviation. 
The quantity $G_{ini}$ asymptotically approaches 1 in all parameter cases. 
Heterogeneity causes $c_v$ to increase across all cases with asymptotically constant behavior as the time leading terms of the numerator and denominator of $c_v$ cancel. Heterogeneity causes $G_{ini}$ to generally increase more rapidly. 
The differences in progressive and regressive $G_{ini}$ are negligible after intermediate times, suggesting that the initial configuration has little effect on its long-term macroscopic dynamics. 
Volatility strongly determines the value of $c_v$ for homogeneous and heterogeneous populations, while only significantly affecting the behavior of $G_{ini}$ in the homogeneous case.

The dynamics of $G_{ini}$ are thus straightforward, but the dependence of $c_v$ on both growth rates and dynamical variance manifests interesting dynamics. To probe the  dynamics of $c_v$, we produce quasi-phase diagrams displaying $\partial_tc_v$ for a continuous spectrum of volatilities over time. 
Fig. \ref{fig:ID}  demonstrates these dynamical regimes for the homogeneous and heterogeneous progressive and regressive cases. 
It shows the magnitude of the initial increase in $c_v$ scales with $\sigma$, as does the magnitude of the eventual decrease. 
The eventual decrease is weaker in regressive populations, and is nonexistent for any volatility value of the progressive population due to the lower progressive growth rate.
Furthermore, they reveal a divergent regime past the critical volatility, where resources decrease while inequality continues to increase. 

These results show that growth rate fluctuations pose a general challenge to any proposed mechanism to address widening inequality that must be addressed independently of volatility in single agent trajectories. Specifically, dynamical balancing of growth rates is required to control long term inequality dynamics. This raises the issue that any method for reducing growth rate fluctuations while preserving growth must include knowledge of the origins of wealth growth rate statistics. Extending the stochastic growth models commonly used
to include environmental opportunities and agent-based learning, in ways analogous to gambling \cite{Kelly} and portfolio theory \cite{hilbert2017more}, naturally produces situations where individual growth rates become distinct and history dependent but also where average growth rates can be maximized and variances minimized over time~\cite{Kelly,bettencourt,cover}. In such dynamical picture, initial correlations will also take on dynamical properties in ways that we stated anticipating here. In recognizing recent research on the effects of heterogeneous and dynamical growth on distributions of wealth ~\cite{Stoj,nelson1991conditional,hsing2005economic,cagetti2008wealth,achdou2017income}, we seek a theoretical framework for wealth dynamics that both complements these phenomenological approaches and incorporates strategic agent behavior in statistical environments. These topics will be presented in future work.  

In summary, we explored the effects of fluctuations in growth rates and wealth statistics in the temporal development of inequality in a population of heterogeneous agents. 
We set up the general problem in the context of multiplicative random growth and derived closed form expressions for how the population inequality changes over time, thus identifying the most important parameters at early and later times.
We found that population variances in growth rates, their correlations with resources, and individual temporal volatility to be the primary drivers of inequality in general situations. The sustained presence of variance in growth rates in a population produces a dominant effect on long-term inequality. While redistribution methods, such as those explored by mean-field or other more complex network exchange models \cite{bouchaud,berman2, garlaschelli} reduce the dynamical impacts of volatility, we have shown that directly addressing growth rate variance in a population is a more fundamental and general requirement for arresting or reversing increases in wealth inequality.

\noindent {\bf Acknowledgements:} We thank Arvind Murugan, Thomas Bourany, and Akhil Ghanta for discussions and comments on the manuscript. This work is supported by the Mansueto Institute for Urban Innovation and the Department of Physics at the University of Chicago and by a National Science Foundation Graduate Research Fellowship (Grant No. DGE 1746045 to JTK).



\bibliographystyle{ieeetr}
\bibliography{refs} 

\appendix
\section{Parameter Covariances}\label{apA}

The time dependent resources of an agent is given by Eq.~\ref{res}

\begin{equation}
    r(t)=r_0\exp{\bigg[\bigg(\eta-\frac{\sigma^2}{2}\bigg)t+\sigma W(t)}\bigg]
\end{equation}

We compute the population effective growth rate over population of size $N$ by separating the  expected growth rate into a mean and covariance term

\begin{eqnarray}
   (\gamma r)_N&=&\frac{1}{N}\sum_{j=1}^N\gamma_jr_j\\ 
&=&\frac{1}{N}\sum_{j=1}^N(\gamma_j-\bar\gamma)r_j+\frac{1}{N}\sum_{j=1}^N\bar\gamma r_j\\ 
&=&\bar \gamma\bar r+\textrm{covar}_N(\gamma_i,r_i)\\
\end{eqnarray}

\noindent where the population effective growth rate can be factored out as $\gamma^\prime=\bar\gamma+\textrm{covar}_N(\gamma_i,\frac{r_i}{\bar r})$.

Assuming $r$ is a lognormal distributed quantity, then so is $r/\bar r$. Define the normal variable $y\equiv\ln r/\bar r\sim \mathcal N(0,\sigma_r)$, then the covariance is equal to $\textrm{covar}_N(\gamma,e^y)$. The first moment is calculated

\begin{equation}
\begin{split}
    \mathrm E[\gamma e^y]&=\int d\gamma dy\gamma e^yP(\gamma,y) \\
    &=\int dye^y\mathrm{E}[\gamma|Y=y]=\int dye^y(\bar\gamma+\rho\frac{\sigma_\gamma}{\sigma_y}y)\\
    &=\bar\gamma\mathrm E[e^y]+\rho\frac{\sigma_\gamma}{\sigma_y}\mathrm E[ye^y]=\bar\gamma e^{\sigma^2_y/2}+\rho\sigma_\gamma\sigma_ye^{\sigma^2_y/2}
\end{split}
\end{equation}

Where the expectation value of $ye^y$ is calculated in ~\cite{mathstack}. With an expectation value defined, the covariance becomes

\begin{equation}
    \begin{split}
        \textrm{Cov}[\gamma,e^y]&=\textrm{E}[\gamma e^\gamma]-\textrm{E}[\gamma]\textrm{E}[e^y]\\
        &=(\bar\gamma+\rho\sigma_\gamma\sigma_y)e^{\sigma^2_y/2}-\bar\gamma e^{\sigma^2_y/2}\\
        &=\rho\sigma_\gamma\sigma_ye^{\sigma^2_y/2}
    \end{split}
\end{equation}

\section{\label{apB}Convolution of distributions}

We compute a joint probability distribution by convolving our dynamical growth distribution from Eq.~\ref{lognorm} with a distribution of initial resources and growth rates.

\begin{equation}\label{B1}
    P(\ln r,t)= \int P(\ln r,t|\gamma, \ln r_0)P(\ln r_0,\gamma) d\ln r_0 d\gamma
\end{equation}

The log resources, $\ln{r_0}$, and growth rates follows a bivariate normal distribution

\begin{equation}
    P(\ln r_0,G)=\mathcal{N}\big(\{\mu_0,G\},K_\Sigma\big)
\end{equation}

for mean initial resources $\mu_0$ and mean growth rate $G$. $\Sigma$ gives the covariance matrix, and will be calculated through the convolution process. The respective variances are given by $\sigma_r,\sigma_G$ with correlation coefficient $\rho$. The growth rate terms can be factored out of the resource integral by completing the square. The integral is then evaluated using Gaussian integral identities, leaving a second round of completing the square, this time for $\mu_0$. The resulting coefficients reduce, leaving the Gaussian  equation

\begin{equation}
\label{convApp}
    P(\ln r,t)= \frac{d \ln r}{\sqrt{2 \pi \Sigma^2}}
    \exp \bigg[ -\frac{(\ln r-\mu_0-G t)^2}{2\Sigma^2}\bigg]
\end{equation}

Where $\Sigma^2=\sigma_r+\sigma^2 t+2\sigma_r\sigma_G\rho+\sigma^2 t^2 $. The covariance matrix is thus deduced as
\begin{equation}
    K_\Sigma=\begin{pmatrix}
\sigma_r^2 & 0 & \sigma_r\sigma_G\rho t \\
0 &  \sigma^2t & 0 \\
\sigma_G\sigma_r\rho t & 0& \sigma_G^2t^2 \\
\end{pmatrix}
\end{equation}




\end{document}